\documentclass[%
 reprint,
 amsmath,amssymb,
 aps,
]{revtex4-1}

\usepackage{graphicx}
\usepackage{dcolumn}
\usepackage{bm}
\usepackage{amsmath}
\usepackage{breqn}

\begin{document}


\title{New Solutions To The Bianchi VIII Wheeler DeWitt Equation And Leading Order Solutions For $\Lambda$ $\ne$ 0 And A Primordial Magnetic Field}
\thanks{A footnote to the article title}%

\author{Daniel Berkowitz}
 \altaffiliation{Physics Department, Yale University.\\ daniel.berkowitz@yale.edu \\ This work is in memory of my parents, Susan Orchan Berkowitz, and Jonathan Mark Berkowitz}

\date{\today}

\begin{abstract}
 New non trivial solutions to the Lorentzian-signature symmetry reduced Bianchi VIII Wheeler DeWitt equation for Hartle Hawking ordering parameters $ \pm  2\sqrt{33} $ are obtained using an Euclidean-signature semi classical method. Using the aforementioned method we define 'excited' states for the Bianchi VIII wave function and construct leading order 'excited' states which appear to be a hybrid of scattering and bound states. Also six new solutions for the $\Lambda \ne 0$ Bianchi VIII Euclidean-signature Hamilton Jacobi equation are presented which can be used to construct semi-classical states corresponding to the Lorentzian signature symmetry reduced Wheeler DeWitt equation. In addition, new solutions to the Euclidean-signature Hamilton Jacobi equation for the Bianchi VIII models are found when $\Lambda \ne 0$ and an aligned primordial magnetic field is present. Furthermore, we find eight new complex solutions to the Euclidean-signature Hamilton Jacobi equation for the case when matter is not present, which can be used to construct additional semi-classical states. We also study the leading order wave functions which result from our aforementioned solutions when they are restricted to the $\beta+$ axis. Prior to presenting these results we explain the Euclidean-signature semi classical method and discuss how to solve the resultant equations this method generates when it is applied to the Lorentzian-signature symmetry reduced Bianchi VIII Wheeler DeWitt equation. The Euclidean-signature semi-classical method used here is applicable to other field theories as well. The work presented throughout this paper further shows the power Eulclidean-signature semi classical methods possess for solving problems in quantum cosmology and solving Lorentzian signature problems in general.

\end{abstract}

\pacs{Valid PACS appear here}
\maketitle


\section{\label{sec:level1}INTRODUCTION}

Over the past 40 years there has been immense progress towards finding a complete theory of quantum gravity\cite{ashtekar1986new,ashtekar1987new,thiemann2006phoenix,maldacena1999large}. One approach to quantum gravity is centered around the Wheeler Dewitt equation\cite{dewitt1967quantum}; a functional differential equation obtained by directly quantizing the ADM\cite{arnowitt1959dynamical} Hamiltonian constraint of general relativity. Strictly speaking due to the operator ordering ambiguity present when quantizing a product of classical variables, there is no unique way of writing the Wheeler DeWitt equation. However for a particular operator ordering the Wheeler DeWitt equation in the total absence of matter can be written as\cite{calcagni2017classical}

\begin{equation}
\begin{aligned}
& \left(\frac{16 \pi G \hbar^{2}}{c^{4} \sqrt{h}} G_{a b c d} \frac{\delta}{\delta h(x)_{a b}} \frac{\delta}{\delta h(x)_{c d}}+\frac{^{(3)}{R} \sqrt{h} c^{4}}{16 \pi G}\right) \Psi=0 \\ &
G_{a b c d}=h_{a c} h_{b d}+h_{a d} h_{b c}-\frac{1}{2}h_{a b} h_{c d},
\end{aligned}
\end{equation}
where $^{(3)}{R}$ is the Ricci scalar curvature for the spatial metric $h_{a b}$, and $\sqrt{h}$ is the 
determinant of $h_{a b}$. The solutions of this equation are known as "wave functions of the universe"\cite{hartle1983wave} and contain all of the information in regards to what possible geometries a particular universe can possess. Solutions to the full functional Wheeler Dewitt equation are incredibly difficult to come by, and thus we don't have a full picture of how Wheeler DeWitt quantum gravity effects cosmological evolution, in the same way we do with classical gravity (general relativity). To obtain a glimpse of how Wheeler DeWitt quantum gravity effects cosmological evolution a minisuperspace approximation is usually employed. It is called an approximation because applying it reduces the degrees of freedom of the classical theory one is trying to quantize. When applied in a cosmological setting it freezes the inhomogeneous degrees of freedom in general relativity, which results in one quantizing a theory of gravity with only finitely as opposed to infinitely many degrees of freedoms. The freezing of inhomogeneous modes results in the loss of quantum effects which originate from those modes in the full theory, thus the equations we obtain cannot capture the full effects quantum gravity has on cosmological evolution. The upside to this approximation though, is that the resulting equation is a partial differential equation as opposed to a functional differential equation, called the symmetry reduced Wheeler DeWitt equation, which is much easier to handle and in certain cases can be solved in closed form. For diagonalized Bianchi A models in general their symmetry reduced Wheeler DeWitt equations are \cite{ryan1975relativistic}
\begin{equation}
\square \Psi-B \frac{\partial \Psi}{\partial \alpha}+U \Psi=0
\end{equation}
where $\square$ is the three-dimensional d'Alembertian in minisuperspace $\left(\alpha,\beta_+,\beta_-\right)$, with signature $(+--)$,  U is a cosmological potential, B can be any real number, and is the Hartle Hawking ordering parameter\cite{hartle1983wave}. This symmetry reduced Wheeler Dewitt equation is derived by inserting a diagonalized Bianchi A metric which has the following form 
\begin{equation}
d s^{2}=-N^{2}d t^{2}+\frac{L^{2}}{6\pi}e^{2 \alpha(t)}\left(e^{2 \beta(t)}\right)_{a b} \omega^{a} \omega^{b} 
\end{equation}
into the Einstein-Hilbert action written in terms of ADM variables \begin{equation}
\mathcal{S}=\frac{c^3}{16\pi G}\int d t d^{3} x N \sqrt{h}\left(K_{a b} K^{a b}-K^{2}+ ^{(3)}{R}-2\Lambda\right);
\end{equation} where $K_{a b}$ are the components of the extrinsic curvature which measure the curvature induced on the Riemannian manifold equipped with spatial metric $h_{a b}$, from the higher dimensional space-time it is embedded in; and $L$ is a constant with dimensions of length which can in principle be absorbed into $\alpha(t)$ as shift in it. After computing the ADM variables in the above action the Lagrnagian density can be found which allows one to construct a Hamiltonian density, which due to the constrained nature of general relativity vanishes. Afterwards the kinetic term in the Hamiltonain constraint will look like \begin{equation}
e^{-3 \alpha}\left(-p_{\alpha}^{2}+p_{+}^{2}+p_{-}^{2}\right)
\end{equation} and can be quantized as follows \begin{equation}
\begin{array}{l}{-e^{-3 \alpha} p_{\alpha}^{2} \longrightarrow \frac{\hbar^{2}}{e^{(3-B) \alpha}} \frac{\partial}{\partial \alpha}\left(e^{-B \alpha} \frac{\partial}{\partial \alpha}\right)} \\ {e^{-3 \alpha} p_{+}^{2} \longrightarrow \frac{-\hbar^{2}}{e^{3 \alpha}} \frac{\partial^{2}}{\partial \beta_{+}^{2}}} \\ {e^{-3 \alpha} p_{-}^{2} \longrightarrow \frac{-\hbar^{2}}{e^{3 \alpha}} \frac{\partial^{2}}{\partial \beta_{-}^{2}}}\end{array}
\end{equation} which results in a equation which can be rearranged into (2).
In the notation introduced by Misner\cite{misner1969quantum}\cite{misner1969mixmaster} $\alpha\left(t\right)$ roughly measures the local scale factor of the spatial surface and $\left(e^{2 \beta(t)}\right)_{i j}$ is $\operatorname{diag}\left(e^{2 \beta\left(t\right)_{+}+2 \sqrt{3} \beta\left(t\right)_{-}}, e^{2 \beta\left(t\right)_{+}-2 \sqrt{3} \beta\left(t\right)_{-}}, e^{-4 \beta\left(t\right)_{+}}\right)$ where $\beta_+$ and $\beta_-$ measure the amount of anisotropy present on the spatial hypersurface. The $\omega^{i}$ terms are one forms defined on the spatial hypersurface of each Bianchi cosmology and obey $ d \omega^{i}=\frac{1}{2} C_{j k}^{i} \omega^{j} \wedge \omega^{k}$ where $C_{j k}^{i}$ are the structure constants of the invariance Lie group associated with each particular class of Bianchi models. For the Bianchi VIII models which is the main focus of this paper the one forms are 
\begin{equation}
\begin{aligned} \omega^{1} &=d x- \sinh (y) d z \\ \omega^{2} &=\cos (x) d y-\sin (x) \cosh (y) d z \\ \omega^{3} &=\sin (x) d y+\cos (x) \cosh (y) d z \end{aligned}. 
\end{equation}

Using the methodology presented in \cite{uggla1995classifying} for writing out cosmological potentials for Bianchi A models with certain matter sources such as a cosmological constant and a primordial magnetic field we will use the Euclidean-signature semi classical method to study the following Lorentizian signature Wheeler DeWitt equation
\begin{equation}
\begin{aligned} 
& \square \Psi-B \frac{\partial \Psi}{\partial \alpha}+U \Psi=0 \\ &
U=\left(f\right) 4 e^{6 \beta_+} \left(e^{6 \beta_+} \sinh ^2\left(2 \sqrt{3}
   \beta_-\right)+\cosh \left(2 \sqrt{3} \beta_-\right)\right) \\& +\frac{2 e^{6 a} \Lambda}{9 \pi }+2b^{2}e^{2\alpha-4\beta_+}+f\\& f=\frac{1}{3} e^{4 \alpha-8 \beta_+}
\end{aligned}
\end{equation}
in units where $G=1,\hspace{1mm} c=1, \hspace{1mm} \hbar=1,\hspace{1mm} \text{and} \hspace{1mm}  L=1$. Also we choose a different normalization for $\Lambda$ than the one presented in \cite{uggla1995classifying}.

This equation (8) can be seen as the analogue of the Schr$\text{\" o}$dinger equation for Wheeler DeWitt Bianchi VIII quantum cosmology. However, it possesses many fundamental differences with the Schr$\text{\" o}$dinger equation which obscures the meaning behind $\psi$. Two notable differences are the absence of any first order time derivative, and the requirement that physically meaningful $\psi$'s must be annihilated by the quantized Hamiltonian constraint $\hat{\mathcal{H}}$, this leads to the problem of time manifesting as
\begin{equation}
\begin{aligned}
& i \hbar \frac{\partial \Psi}{\partial t}=N \hat{\mathcal{H}}_{\perp} \Psi \\ &
\frac{\partial \Psi}{\partial t}=0.
\end{aligned}
\end{equation}

A way around this for our purposes is to denote one of the Misner variables to be our clock. A good clock increases monotonically. Out of the variables we can choose from, $\alpha$ which corresponds to the scale factor of a Bianchi VIII universe is the best candidate for our clock and will be for practical purposes our "time" \cite{dewitt1967quantum} parameter. However, due to the immense difficulty in constructing a dynamical unitary operator from the symmetry reduced Wheeler DeWitt equation, $|\psi\left(\alpha,\beta_+,\beta_-\right)|^{2}$ is not conserved in $\alpha$, thus we cannot assign a simple probabilistic interpretation to our wave function. Solutions to the symmetry reduced Wheeler DeWitt equation can be interpreted through more elaborate means \cite{griffiths1984consistent}\cite{bohm1952suggested}. Furthermore our solutions can be expressed as a complex valued wave function, thus allowing one to analyze them using a Klein-Gordon current \cite{vilenkin1989interpretation} \cite{mostafazadeh2004quantum}
\begin{equation}
\mathcal{J}=\frac{i}{2}\left(\psi^{*} \nabla \psi-\psi \nabla \psi^{*}\right).
\end{equation}
A quantitative analysis of our mathematical results will be left to future works. 
This paper will have the following structure. First, we will explain the  Euclidean-signature semi classical method developed by Moncrief et al. \cite{marini2019euclidean} and discuss how to solve the resultant equations this method generates when it is applied to the symmetry reduced Bianchi VIII Wheeler DeWitt equation. We will then use the solutions to the resultant equations to construct new solutions for the symmetry reduced Bianchi VIII Wheeler DeWitt equation for the case where $\Lambda=0$ for two particular Hartle Hawking parameters. We will then define 'excited' states and construct semi-classical states when $\Lambda=0$. Afterward, we will present a plethora of other solutions to the Euclidean-signature Hamilton Jacobi equation for the case when $\Lambda$ $\ne$ 0, when $\Lambda=0$, and when both a cosmological constant and a primordial field are present, which on their own can be used to construct a variety of leading order solutions. Afterwards we will analyze the leading order solutions constructed from the aforementioned solutions to the Euclidean-signature Hamiliton Jacobi equation for the case when we restrict ourselves to the $\beta_+$ axis. Finally we will give closing remarks. It should be noted that past this point all references to the Wheeler DeWitt equation unless otherwise stated refer to the symmetry reduced Bianchi VIII Wheeler DeWitt equation. Also the terms leading order solution and semi-classical state will be used interchangeably. 
\section{\label{sec:level1}The Euclidean-signature semi classical method} 
Our outline of this method will follow closely \cite{moncrief2014euclidean}. The method outlined in this section and its resultant equations can in principle be used to construct solutions (closed form and asymptotic) to a wide class of quantum cosmological models such as all of the Bianchi A models and their corresponding locally rotationally symmetric (LRS) models expressed in Misner variables. 
 
The first step we will take in solving the Wheeler DeWitt equation is to introduce the ansatz
 \begin{equation}
\stackrel{(0)}{\Psi}_{\hbar}=e^{-S_{\hbar} / \hbar}
\end{equation}
where $S_{\hbar}$ is a function of $\left(\alpha,\beta_+,\beta_-\right)$ We will rescale $S_{\hbar}$ in the following way  
\begin{equation}
\mathcal{S}_{\hbar} :=\frac{G}{c^{3} L^{2}} S_{\hbar}
\end{equation}
where $\mathcal{S}_{\hbar}$ is dimensionless and admits the following power series in terms of this dimensionless parameter
\begin{equation}
X :=\frac{L_{\text { Planck }}^{2}}{L^{2}}=\frac{G \hbar}{c^{3} L^{2}}.
\end{equation}

The series is given by 
\begin{equation}
\mathcal{S}_{\hbar}=\mathcal{S}_{(0)}+X \mathcal{S}_{(1)}+\frac{X^{2}}{2 !} \mathcal{S}_{(2)}+\cdots+\frac{X^{k}}{k !} \mathcal{S}_{(k)}+\cdots
\end{equation},
and as a result our initial ansatz now takes the following form 
\begin{equation}
\stackrel{(0)}{\Psi}_{\hbar}=e^{-\frac{1}{X} \mathcal{S}_{(0)}-\mathcal{S}_{(1)}-\frac{X}{2 !} \mathcal{S}_{(2)}-\cdots}
\end{equation}.

Substituting this ansatz into the Wheeler-DeWitt equation and
requiring satisfaction, order-by-order in powers of X leads immediately to the sequence of equations

\begin{equation}
\begin{aligned}
&{\left(\frac{\partial \mathcal{S}_{(0)}}{\partial \alpha}\right)^{2}-\left(\frac{\partial \mathcal{S}_{(0)}}{\partial \beta_{+}}\right)^{2}-\left(\frac{\partial \mathcal{S}_{(0)}}{\partial \beta_{-}}\right)^{2}}+U=0
\end{aligned}
\end{equation}
\begin{equation}
\begin{aligned}
& 2\left[\frac{\partial \mathcal{S}_{(0)}}{\partial \alpha} \frac{\partial \mathcal{S}_{(1)}}{\partial \alpha}-\frac{\partial \mathcal{S}_{(0)}}{\partial \beta_{+}} \frac{\partial \mathcal{S}_{(1)}}{\partial \beta_{+}}-\frac{\partial \mathcal{S}_{(0)}}{\partial \beta_{-}} \frac{\partial \mathcal{S}_{(1)}}{\partial \beta_{-}}\right] \\ & +B \frac{\partial \mathcal{S}_{(0)}}{\partial \alpha}-\frac{\partial^{2} \mathcal{S}_{(0)}}{\partial \alpha^{2}}+\frac{\partial^{2} \mathcal{S}_{(0)}}{\partial \beta_{+}^{2}}+\frac{\partial^{2} \mathcal{S}_{(0)}}{\partial \beta_{-}^{2}}=0,
\end{aligned}
\end{equation},
\begin{equation}
\begin{aligned}
& 2\left[\frac{\partial \mathcal{S}_{(0)}}{\partial \alpha} \frac{\partial \mathcal{S}_{(k)}}{\partial \alpha}-\frac{\partial \mathcal{S}_{(0)}}{\partial \beta_{+}} \frac{\partial \mathcal{S}_{(k)}}{\partial \beta_{+}}-\frac{\partial \mathcal{S}_{(0)}}{\partial \beta_{-}} \frac{\partial \mathcal{S}_{(k)}}{\partial \beta_{-}}\right] \\ & {+k\left[B \frac{\partial \mathcal{S}_{(k-1)}}{\partial \alpha}-\frac{\partial^{2} \mathcal{S}_{(k-1)}}{\partial \alpha^{2}}+\frac{\partial^{2} \mathcal{S}_{(k-1)}}{\partial \beta_{+}^{2}}+\frac{\partial^{2} \mathcal{S}_{(k-1)}}{\partial \beta_{-}^{2}}\right]} \\ & + \sum_{\ell=1}^{k-1} \frac{k !}{\ell !(k-\ell) !}\Biggr(\frac{\partial \mathcal{S}_{(\ell)}}{\partial \alpha} \frac{\partial \mathcal{S}_{(k-\ell)}}{\partial \alpha}-\frac{\partial \mathcal{S}_{(\ell)}}{\partial \beta_{+}} \frac{\partial \mathcal{S}_{(k-\ell)}}{\partial \beta_{+}} \\& - \frac{\partial \mathcal{S}_{(\ell)}}{\partial \beta_{-}} \frac{\partial \mathcal{S}_{(k-\ell)}}{\partial \beta_{-}}\Biggl) =0
\end{aligned}
\end{equation}

We will refer to $\mathcal{S}_{(0)}$ in our Wheeler DeWitt wave functions as the leading order term, which can be used to construct a semi-classical approximate solution to the Lorentzian signature Wheeler DeWitt equation, and call $\mathcal{S}_{(1)}$ the first order term. One reason why the $\mathcal{S}_{(0)}$ term can also be called the semi-classical term, beyond the fact that it satisfies the classical Euclidean-signature Hamilton Jacobi equation is that the equation it satisfies is independent of the ambiguities which result from operator ordering. The $\mathcal{S}_{(1)}$ term can also be viewed as our first quantum correction, with the other $\mathcal{S}_{(k)}$ terms being the additional higher order quantum corrections. This is reflected in the fact that the higher order transport equations depend on the operator ordering used in defining the Wheeler Dewitt equation, which is an artifact of quantization. Our closed form wave functions will not require terms beyond first order.

If one can find a solution to the $\mathcal{S}_{(1)}$ equation which allows the $\mathcal{S}_{(2)}$ equation to be satisfied by zero, then one can write down the following as a solution to the Wheeler DeWitt equation for either a particular value of the Hartle-Hawking ordering parameter, or for an arbitrary ordering parameter depending on the properties of the $\mathcal{S}_{(1)}$ which is found.   

\begin{equation}
\stackrel{(0)}{\Psi}_{\hbar}=e^{-\frac{1}{X} \mathcal{S}_{(0)}-\mathcal{S}_{(1)}}
\end{equation}.

This can be easily shown. Lets take $\mathcal{S}_{(0)}$ and $\mathcal{S}_{(1)}$ as arbitrary known functions which allow the $\mathcal{S}_{(2)}$ transport equation to be satisfied by zero, then the $k=3$ transport equation can be expressed as 
\begin{equation}
{2\left[\frac{\partial \mathcal{S}_{(0)}}{\partial \alpha} \frac{\partial \mathcal{S}_{(3)}}{\partial \alpha}-\frac{\partial \mathcal{S}_{(0)}}{\partial \beta_{+}} \frac{\partial \mathcal{S}_{(3)}}{\partial \beta_{+}}-\frac{\partial \mathcal{S}_{(0)}}{\partial \beta_{-}} \frac{\partial \mathcal{S}_{(3)}}{\partial \beta_{-}}\right]}=0
\end{equation}
which is clearly satisfied by $\mathcal{S}_{(3)}$=0. The $\mathcal{S}_{(4)}$ equation can be written in the same form as (20) and one of its solution is 0 as well, thus resulting in the $\mathcal{S}_{(5)}$ equation possessing the same form as (20). One can easily convince oneself that this pattern continues for all of the $k\geq 3$ $\mathcal{S}_{(k)}$ transport equations as long as the solution of the $\mathcal{S}_{(k-1)}$ transport equation is chosen to be 0. Thus in some situations an $\mathcal{S}_{(1)}$ exists which allows one to set the solutions to all of the higher order transport equations to zero, which results in the infinite sequence of transport equations generated by our ansatz to truncate to a finite sequence of equations whose solutions allow us to construct a closed form wave function satisfying the Wheeler DeWitt equation. Not all solutions to the $\mathcal{S}_{(1)}$ transport equation will allow the $\mathcal{S}_{(2)}$ transport equation to be satisfied by zero; however in our case, we were able to find $\mathcal{S}_{(1)}$'s which cause the $\mathcal{S}_{(2)}$ transport equation to be satisfied by zero, thus allowing one to set all of the solutions to the higher order transport equations to zero as shown above. This will enable us to construct closed form solutions to the Lorentzian signature Bianchi VIII Wheeler Dewitt equation for particular Hartle Hawking ordering parameters.   

To calculate 'excited' states we introduce the following ansatz. 
\begin{equation}
{\Psi}_{\hbar}={\phi}_{\hbar} e^{-S_{\hbar} / \hbar}
\end{equation}
where $$
S_{\hbar}=\frac{c^{3} L^{2}}{G} \mathcal{S}_{\hbar}=\frac{c^{3} L^{2}}{G}\left(\mathcal{S}_{(0)}+X \mathcal{S}_{(1)}+\frac{X^{2}}{2 !} \mathcal{S}_{(2)}+\cdots\right)
$$
is the same series expansion as before and ${\phi}_{\hbar}$ can be expressed as the following series 
\begin{equation}
{\phi_{\hbar}=\phi_{(0)}+X \phi_{(1)}+\frac{X^{2}}{2 !} \phi_{(2)}+\cdots+\frac{X^{k(*)}}{k !} \phi_{(k)}+\cdots}
\end{equation}
with X being the same dimensionless quantity as before. 
Inserting $\left(21\right)$ with the expansions given by $\left(14\right)$ and $\left(22\right)$ into the Wheeler DeWitt equation and by matching equations in powers of X leads to the following sequence of equations. 
\begin{equation}
-\frac{\partial \phi_{(0)}}{\partial \alpha} \frac{\partial \mathcal{S}_{(0)}}{\partial \alpha}+\frac{\partial \phi_{(0)}}{\partial \beta_{+}} \frac{\partial \mathcal{S}_{(0)}}{\partial \beta_{+}}+\frac{\partial \phi_{(0)}}{\partial \beta_{-}} \frac{\partial \mathcal{S}_{(0)}}{\partial \beta_{-}}=0,
\end{equation},
\begin{equation}
\begin{aligned}
&{-\frac{\partial \phi_{(1)}}{\partial \alpha} \frac{\partial \mathcal{S}_{(0)}}{\partial \alpha}+\frac{\partial \phi_{(1)}}{\partial \beta_{+}} \frac{\partial \mathcal{S}_{(0)}}{\partial \beta_{+}}+\frac{\partial \phi_{(1)}}{\partial \beta_{-}} \frac{\partial \mathcal{S}_{(0)}}{\partial \beta_{-}}} \\ & {+\left(-\frac{\partial \phi_{(0)}}{\partial \alpha} \frac{\partial \mathcal{S}_{(1)}}{\partial \alpha}+\frac{\partial \phi_{(0)}}{\partial \beta_{+}} \frac{\partial \mathcal{S}_{(1)}}{\partial \beta_{+}}+\frac{\partial \phi_{(0)}}{\partial \beta_{-}} \frac{\partial \mathcal{S}_{(1)}}{\partial \beta_{-}}\right)} \\ & {+\frac{1}{2}\left(-B \frac{\partial \phi_{(0)}}{\partial \alpha}+\frac{\partial^{2} \phi_{(0)}}{\partial \alpha^{2}}-\frac{\partial^{2} \phi_{(0)}}{\partial \beta_{+}^{2}}-\frac{\partial^{2} \phi_{(0)}}{\partial \beta_{-}^{2}}\right)=0,}
\end{aligned}
\end{equation}
\begin{equation}
\begin{aligned}
& -\frac{\partial \phi_{(k)}}{\partial \alpha} \frac{\partial \mathcal{S}_{(0)}}{\partial \alpha}+\frac{\partial \phi_{(k)}}{\partial \beta_{+}} \frac{\partial \mathcal{S}_{(0)}}{\partial \beta_{+}}+\frac{\partial \phi_{(k)}}{\partial \beta_{-}} \frac{\partial \mathcal{S}_{(0)}}{\partial \beta_{-}} \\ & 
+k\Biggr(-\frac{\partial \phi_{(k-1)}}{\partial \alpha} \frac{\partial \mathcal{S}_{(1)}}{\partial \alpha}+\frac{\partial \mathcal{S}_{(1)}}{\partial \beta_{+}} \frac{\partial \mathcal{S}_{(1)}}{\partial \beta_{+}}+\frac{\partial \phi_{(k-1)}^{(*)}}{\partial \beta_{-}} \frac{\partial \mathcal{S}_{(1)}}{\partial \beta_{-}}\Biggr) \\ & 
+\frac{k}{2}\Biggr(-B \frac{\partial \phi_{(k-1)}}{\partial \alpha}+\frac{\partial^{2} \phi_{(k-1)}^{(*)}}{\partial \alpha^{2}}-\frac{\partial^{2} \phi_{(k-1)}}{\partial \beta_{+}^{2}}-\frac{\partial^{2} \phi_{(k-1)}}{\partial \beta_{-}^{2}}\Biggr)\\ & -
\sum_{\ell=2}^{k} \frac{k !}{\ell !(k-\ell) !}\Biggr( \frac{\partial \phi_{(k-\ell)}}{\partial \alpha} \frac{\partial \mathcal{S}_{(\ell)}}{\partial \alpha}-\frac{\partial \phi_{(k-\ell)}}{\partial \beta_{+}} \frac{\partial \mathcal{S}_{(\ell)}}{\partial \beta_{+}} - \\ &   \frac{\partial \phi_{(k-\ell)}}{\partial \beta_{-}} \frac{\partial \mathcal{S}_{(\ell)}}{\partial \beta_{-}}\Biggr) =0.
\end{aligned}
\end{equation}

It can be seen from computing $\frac{d\phi_{(0)}\left(\alpha,\beta_+,\beta_-\right)}{dt}=\dot{\alpha}\frac{\partial \phi_{(0)}}{\partial \alpha}+\dot{\beta_+}\frac{\partial \phi_{(0)}}{\partial \beta_+}+\dot{\beta_-}\frac{\partial \phi_{(0)}}{\partial \beta_-}$, and inserting $\left(4.9, \hspace{1 mm} 4.18-4.20\right)$ from \cite{moncrief2014euclidean} that $\phi_{(0)}$ is a conserved quantity under the flow of $S_{0}$. This means any function $F\left(\phi_{(0)}\right)$ is also a solution of equation $\left(26\right)$. Wave functions constructed from these functions of $\phi_{0}$ are only physical if they are smooth and globally defined. Beyond the semi-classical limit, if smooth globally defined solutions can be proven to exist for the higher order $\phi$ transport equations then one can construct a family of 'excited' states and take a superposition of them like in ordinary quantum mechanics.

Our 'excited' states ${\Psi}_{\hbar}={\phi}_{\hbar} e^{-S_{\hbar} / \hbar}$ qualitatively possess the same form as excited states for the quantum harmonic oscillator $ \psi_{n}(x)=H_{n}\left(\sqrt{\frac{m \omega}{\hbar}} x\right)\frac{1}{\sqrt{2^{n} n !}}\left(\frac{m \omega}{\pi \hbar}\right)^{1 / 4}  e^{-\frac{m \omega x^{2}}{2 \hbar}}$, where $H_n$ are the Hermite polynomials in which n is a positive integer which specifies its form. Because the solutions of the $\phi_{(0)}$ equation are quantities conserved along the flow generated by $\mathcal{S}_{(0)}$, any multiple $\phi^{n}_{(0)}$ also satisfies equation $\left(26\right)$. On purely physical grounds the amount of numbers required to specify an 'excited' state equals the number of excitable degrees of freedom present. For Bianchi A models with non dynamical matter sources that amounts to two numbers corresponding to the two anistropic degrees of freedoms. As a result our $\phi_{(0)}$ which distinguishes our 'excited' states from 'ground' states has the following form $ \prod_{i = 1}^{n} f^{m_{i}}\left(\alpha,\beta_+,\beta_-\right)_{i}$; where $f\left(\alpha,\beta_+,\beta_-\right)_{i}$ are independent conserved quantities satisfying equation $\left(26\right)$ which are raised to the power $m_{i}$, and n is the number of excitable degrees of freedom. If all of the $f\left(\alpha,\beta_+,\beta_-\right)_{i}$'s vanish at some point or points in minisuperspace then to ensure that our wave function is smooth and globally defined we must restrict $m_{i}$ to be positive integers which results in our 'excited' states being 'bound' states just like the quantum harmonic oscillator. This discretization of the quantities that are used to denote our 'excited' states is the mathematical manifestation of quantization one would expect excited states to possess. If none of our conserved quantities $f\left(\alpha,\beta_+,\beta_-\right)_{i}$ vanish in minisuperspace than our 'excited' states are 'scattering' states akin to the quantum free particle and $m_{i}$ can be any real or complex number. It is also possible that only some $f\left(\alpha,\beta_+,\beta_-\right)_{i}$'s vanish while the other do not, in this case our 'excited' states are a hybrid of 'bound' and 'scattering' states which means some degrees of freedoms are 'bound' while other are 'scattering'. This is the case for the quantum Bianchi VIII models we will analyze in this paper. Additional information for why we can call the above 'excited' states despite them being solutions to an equation which does not have the same form as the Schr$\text{\" o}$dinger equation can be found in \cite{moncrief2014euclidean}. In what follows we will set $L=1$, $c=1$, $G=1$ and $\hbar=1$.

\section{\label{sec:level1}Solutions For Wheeler DeWitt Equation With B=$\pm $ $ 2\sqrt{33} $}

The leading order terms for all of our solutions in this section were first computed by  O. Obregón, J. and Socorro \cite{obregon1996psi} and are the following

\begin{equation}
\mathcal{S}_{(0)}:= \pm \frac{1}{6} e^{2 \alpha-4 \beta_+} \left(2 e^{6 \beta_+} \cosh \left(2 \sqrt{3} \beta_-\right)-1\right).
\end{equation}

Starting with the above pair of solutions to the Bianchi VIII Euclidean-signature Hamilton Jacobi equation, O. Obregón, J. and Socorro \cite{obregon1996psi} found the following two solutions to the Bianchi VIII Wheeler DeWitt equation for Hartle Hawking parameters $B=6$ and $B=-6$ respectively 
\begin{equation}
\psi_{\mp}:= e^{6\alpha \mp \mathcal{S}_{(0)}}
\end{equation}
\begin{equation}
\psi_{\mp}:= e^{\mp \mathcal{S}_{(0)}}.
\end{equation}
In terms of the formalism presented in the previous section these two solutions have the following $\mathcal{S}_{(1)}$ term or quantum correction
\begin{equation}
\mathcal{S}_{(1)}:=-\frac{1}{2} (\text{B}+6)\alpha
\end{equation}
which at ordering parameters $B=\pm6$ cause the source term of the $\mathcal{S}_{(2)}$ equation to vanish, allowing one to construct closed form solutions $\stackrel{(0)}{\Psi}_{\hbar}=e^{\mp\frac{1}{X} \mathcal{S}_{(0)}-\mathcal{S}_{(1)}}$ to the Wheeler Dewitt equation. To construct new solutions we need to find a different $\mathcal{S}_{(1)}$. The $\mathcal{S}_{(1)}$ equation we need to solve is the following
\begin{equation}
\begin{aligned}
& 2 e^{6 \beta_+} \cosh \left(2 \sqrt{3} \beta_-\right) \left(2 \frac{\partial \mathcal{S}_{(1)}}{\partial \alpha}-2 \frac{\partial \mathcal{S}_{(1)}}{\partial \beta_+}+\text{B}+6\right)\\&-2 \frac{\partial \mathcal{S}_{(1)}}{\partial \alpha}-  4 \sqrt{3}
   e^{6 \beta_+} \frac{\partial \mathcal{S}_{(1)}}{\partial \beta_-} \sinh \left(2 \sqrt{3} \beta_-\right)-4 \frac{\partial \mathcal{S}_{(1)}}{\partial \beta_+}\\&-\text{B}-6=0.
\end{aligned}
\end{equation}

Because there are no explicit functions of $\alpha$ in our transport equation we can seek solutions which have the form $\mathcal{S}_{(1)}= x1\alpha +f\left(\beta_+,\beta_-\right)$, where x1 is an arbitrary real or complex number. Inserting this into our transport equation yields the following

\begin{equation}
\begin{aligned}
& 2 e^{6 \beta_+} \cosh \left(2 \sqrt{3} \beta_-\right) \left(-2 \frac{\partial f_{(1)}}{\partial \beta_+}
+\text{B}+2 \text{x1}+6\right)-4 \frac{\partial f_{(1)}}{\partial \beta_+}
\\ &-4 \sqrt{3} e^{6 \beta_+} \sinh \left(2
   \sqrt{3} \beta_-\right) \frac{\partial f_{(1)}}{\partial \beta_-}
-\text{B}-2 \text{x1}-6=0.
\end{aligned}
\end{equation}

This is a non homogeneous transport equation with variable coefficients. This can be solved using a computer algebraic systems such as Mathematica or by applying a change of variables to turn this equation into a homogeneous transport equation and applying standard textbook techniques. A solution to this equation is 

\begin{equation}
\begin{aligned}
f:=\frac{1}{8} (\text{B}+2 \text{x1}+6) \left(\log \left(\sinh \left(2 \sqrt{3} \beta_-\right)\right)-2\beta_+\right)
\end{aligned}
\end{equation}
which allows us to construct the following family of solutions to the $\mathcal{S}_{(1)}$ transport equation
\begin{equation}
\begin{aligned}
& \mathcal{S}_{(1)}:=  x1\alpha+f \\ & f:=\frac{1}{8} (\text{B}+2 \text{x1}+6) \left(\log \left(\sinh \left(2 \sqrt{3} \beta_-\right)\right)-2\beta_+\right).
\end{aligned}
\end{equation}
The previous $\mathcal{S}_{(1)}$'s (29) which correspond to the closed form solutions found by O. Obregón, and J. Socorro are a subset of (33) as can be seen by setting $x1=-\frac{1}{2}\left(6+B\right)$. When this more general solution is inserted into the $\mathcal{S}_{(2)}$ transport equation its source term vanishes when $\left(x1=0,B=-6\right)$, $\left(x1=-6,B=6\right)$, and
$\left(x1=-7\pm\sqrt{33},B=\mp2\sqrt{33}\right)$. 
The closed form solutions corresponding to $\left(x1=-7+\sqrt{33},B=-2\sqrt{33}\right)$ and $\left(x1=-7-\sqrt{33},B=2\sqrt{33}\right)$ are our new solutions, and they take the following forms respectively 
\begin{equation}
\psi_{\mp}:= e^{\left(7-\sqrt{33}\right) \alpha-2 \beta_+\mp \mathcal{S}_{(0)}}\sinh \left(2 \sqrt{3} \beta_-\right)
\end{equation}
\begin{equation}
\psi_{\mp}:= e^{\left(7+\sqrt{33}\right) \alpha-2 \beta_+-\mp \mathcal{S}_{(0)}}\sinh \left(2 \sqrt{3} \beta_-\right).
\end{equation}

From now on we will only analyze solutions corresponding to the pair $\left(x1=-7+\sqrt{33},B=-2\sqrt{33}\right)$ because as it can be seen from the above equations, qualitatively the behavior of both of these solutions are practically identical. What we shall do below can be exactly duplicated for the ordering parameter $2\sqrt{33}$

As one can already see our 'ground' state solutions $\psi_{\pm 1}$ blur the line between 'ground' and 'excited' states because they has the following form
$\Psi_{\hbar}=\phi_{\hbar} e^{-S_{\hbar} / \hbar}$, where $\phi_{\hbar}$= $\sinh \left(2 \sqrt{3} \beta_-\right)$, despite originating from solving the 'ground' state sequence of transport equations. This is because the homogeneous form of the $\mathcal{S}_{(1)}$ transport equation is identical to the $\phi_{(0)}$ 'exicted' state transport equation. Thus, one can begin constructing 'excited' states from the $\mathcal{S}_{(1)}$ equation by adding the homogeneous solution to its inhomogeneous solution. As a result one can be tempted to interpret 'excited' states in Wheeler Dewitt quantum cosmology as semi-classical 'ground' states with additional quantum corrections attached to them, assuming the term 'ground' state is appropriate at all to begin with. 

This ambiguity is a result of us not having a rigorous understanding of what constitutes 'ground' states as opposed to 'excited' states in quantum cosmology yet. Ultimately to differentiate between 'ground' and 'excited' states we will appeal to their qualitative features such as having multiple local max/mins or peaks, and their mathematical manifestation of quantization or discreteness. 

To construct additional closed form solutions we can solve equation (23) which will give us the homogeneous solutions to  (17) 
\begin{equation}
\begin{aligned}
&-2 e^{6 \beta_+} \cosh \left(2 \sqrt{3} b\right) (\frac{\partial \phi}{\partial \alpha}-\frac{\partial \phi}{\partial \beta_+})+\frac{\partial \phi}{\partial \alpha}\\&+2 \sqrt{3} e^{6 \beta_+} \frac{\partial \phi}{\partial \beta_-} \sinh
   \left(2 \sqrt{3} \beta_-\right)+2 \frac{\partial \phi}{\partial \beta_+}=0.
\end{aligned}
\end{equation}

The solutions of this equation are functions which are conserved along the flow generated by $\mathcal{S}_{(0)}$. An easy way to find solutions to this equation is to modify the known solutions to the Bianchi IX analogue of this equation(36) which were computed by Joseph Bae \cite{bae2015mixmaster}, and are 
\begin{equation}
\phi^{1}_{IX,0}:=\frac{1}{6} e^{4 \alpha-2 \beta_{+}}\left(e^{6 \beta_{+}}-\cosh \left(2 \sqrt{3} \beta_{-}\right)\right)
\end{equation}
\begin{equation}
\phi^{2}_{IX,0} :=\frac{1}{2 \sqrt{3}} e^{4 \alpha-2 \beta_{+}} \sinh \left(2 \sqrt{3} \beta_{-}\right),
\end{equation}
where the superscripts delineate the two conserved quantities. 

The modification we will apply to the above solutions to the Bianchi IX $\phi_{0}$ equation will be determined by comparing the $\mathcal{S}_{(0)}$ terms of the diagonalized Bianchi VIII models to the diagonalized Bianchi IX models. If we compare $\pm \frac{1}{6} e^{2 \alpha-4 \beta_+} \left(2 e^{6 \beta_+} \cosh \left(2 \sqrt{3} \beta_-\right)-1\right)$ to $\pm \frac{1}{6} e^{2 \alpha-4 \beta_+} \left(2 e^{6 \beta_+} \cosh \left(2 \sqrt{3} \beta_-\right)+1\right)$ we see that mathematically the only difference between these two expressions is that they share opposite signs for their $e^{-4 \beta_+}$ term. Thus we can make an educated guess, and insert the following pair of expressions into equation (36) 
\begin{equation}
\phi^{1}_{0} :=\frac{1}{6} e^{4 \alpha-2 \beta_{+}}\left(e^{6 \beta_{+}}+\cosh \left(2 \sqrt{3} \beta_{-}\right)\right)
\end{equation}
\begin{equation}
\phi^{2}_{0} :=\frac{1}{2 \sqrt{3}} e^{4 \alpha-2 \beta_{+}} \sinh \left(2 \sqrt{3} \beta_{-}\right).
\end{equation}
As the reader can verify this pair satisfies (36) and is conserved along the flow generated (26). We will now denote our two independent conserved quantities as 
\begin{equation}
C:=\frac{1}{6} e^{4 \alpha-2 \beta_{+}}\left(e^{6 \beta_{+}}+\cosh \left(2 \sqrt{3} \beta_{-}\right)\right)
\end{equation}
\begin{equation}
S:=\frac{1}{2 \sqrt{3}} e^{4 \alpha-2 \beta_{+}} \sinh \left(2 \sqrt{3} \beta_{-}\right)
\end{equation}
and take advantage of the fact that any function of a conserved quantity is also a conserved quantity by expressing our  $\phi_{0}$ as 
\begin{equation}
\phi_{0}= S^{m1} C^{m2}.
\end{equation}
The reader should keep in mind that we could have chosen any function of our aforementioned conserved quantities to be our $\phi_{0}$. We choose this particular form for our $\phi_{0}$ so that it can conform to the ans$\text{\" a}$tze which are given in \cite{moncrief2014euclidean} \cite{berkowitz2019new}.

The parameters $\left(m_1, m_2\right)$ can plausibly be interpreted as graviton excitation numbers\cite{bae2014quantizing} for the ultra long wavelength gravitational wave modes embodied in the $\left(\beta_+,\beta_-\right)$ anisotropic degrees of freedom. In the $\beta$ plane $C_{(0)}$ does not vanish while $ S_{(0)}$ does. This means in order to have globally defined wave functions we must restrict $m1$ to be positive integers only while $m2$ can be any real or complex number. This is very peculiar, if we go back to our chosen Bianchi IX conserved quantities (37) and (38), we that they both vanish in the $\beta$ plane, which means that $m1$ and $m2$ are forced to be positive integers to assure that 'excited' states are globally defined and smooth. This quantization of both $m1$ and $m2$ due to the requirement of having a globally defined wave function is the main reason we call states of the form (21) 'excited' states because they are discretized in the same manner as bound states in ordinary quantum mechanics, which are denoted by positive integers $n_{i}$. However, for the Bianchi VIII case we have one conserved quantity which does not vanish in the $\beta$ plane, while another does vanish, this means our 'excited' states as they are currently formulated are some hybrid of a scattering state and a bound state. We will move forward with our analysis while acknowledging that there are some profound differences between 'excited' states of the Bianchi IX models and the Bianchi VIII models. 
We will now use our conserved quantity to construct 'excited' states which have the same form as (21) 
\begin{equation}
\Psi_{\mp}:= S^{m1} C^{m2}e^{\frac{1}{2} \alpha (B+6)\mp\mathcal{S}_{(0)}}
\end{equation}

where we used the $\mathcal{S}_{(1)}$ in equation (29).

By inserting (44) into the Wheeler DeWitt equation we see that for ordering parameters $\pm 2\sqrt{33}$ it is satisfied when $m_1=0$ and $m_2=1$, and vice versa. This results in more closed form solutions, one of which for ordering parameter $ -2\sqrt{33} $ is given below
\begin{equation}
\Psi_{\mp}:= e^{\left(7-\sqrt{33}\right) \alpha-2 \beta_+\mp \mathcal{S}_{(0)}}\left(e^{6 \beta_{+}}+\cosh \left(2 \sqrt{3} \beta_{-}\right)\right).
\end{equation}

Because the Wheeler DeWitt equation is linear we can construct additional closed form solutions by taking linear combinations of the above solutions. Doing so results in the following family of exact solutions 
\begin{equation}
\Psi_:=a_{1}\psi_{-}+a_{2}\psi_{+}+a_{3}\Psi_{-}+a_{4}\Psi_{+}
\end{equation}
where $a_{i}$ can be any arbitrary real or complex number. 

Before we move forward, we have to address the pathologies present in our Bianchi VIII wave functions. When (26) is exponentiated, it results in a wave function which runs off to infinity when $\beta_+$ grows without bound in either direction depending on the sign of (26). This results in a wave function which isn't normalizable when it is integrated over all of $\beta$ space at a fixed $\alpha$, which is in contrast to the Bianchi IX wave functions computed in \cite{moncrief1991amplitude} \cite{bae2015mixmaster}. Those wave functions are normalizable at fixed values of $\alpha$ which opens them up to be admittingly naively interpreted in terms of how likely a Bianchi IX universe is to have a given level of anisotropy dictated by the $\beta$ variables at a particular value of $\alpha$. The plots we will show are in a region of the $\beta$ plane where our wave functions look like 'excited' states for a particular range of $\alpha$.

If we only consider the semi classical limit  

\begin{equation}
\Psi_{\mp}:= S^{m1} C^{m2}e^{\mp\mathcal{S}_{(0)}}
\end{equation}

We find an interesting 'excited' state which illustrates the previously mentioned points about ambiguity in what constitutes an 'excited state. 

If we choose $m_1=2$ and $m_2=-4$, and plot $\Psi^{2}_{-}$ for $\alpha=-3$ we obtain a wave function which has some qualitative features of an 'excited' state. However, if we plot the same wave function for $\alpha=0$ we see that those 'excited' state qualitative features are no longer present, and the interpretation of this state become difficult to ascertain. This is displayed in figure 1 and 2. 

\begin{figure}[!ht]
\begin{minipage}[c]{0.4\linewidth}
\includegraphics[scale=.09]{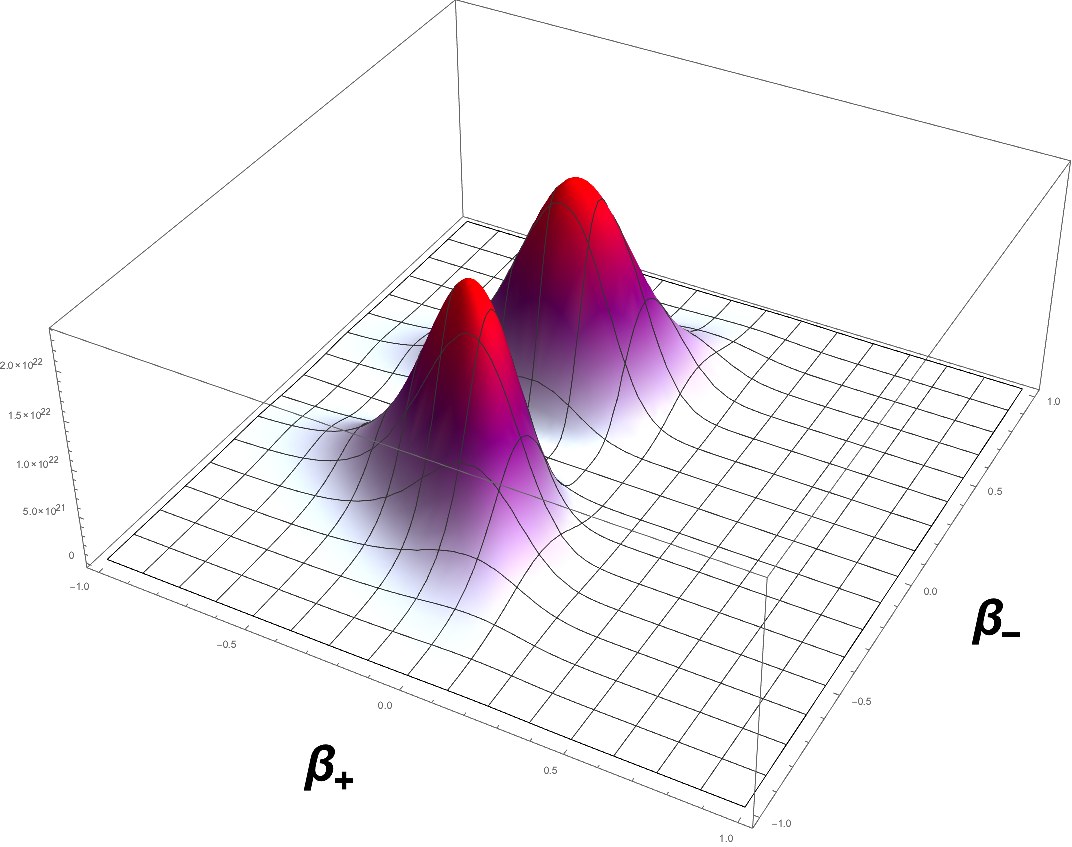}
\caption{$\alpha=-3$}
\end{minipage}

\begin{minipage}[c]{0.4\linewidth}
\includegraphics[scale=.09]{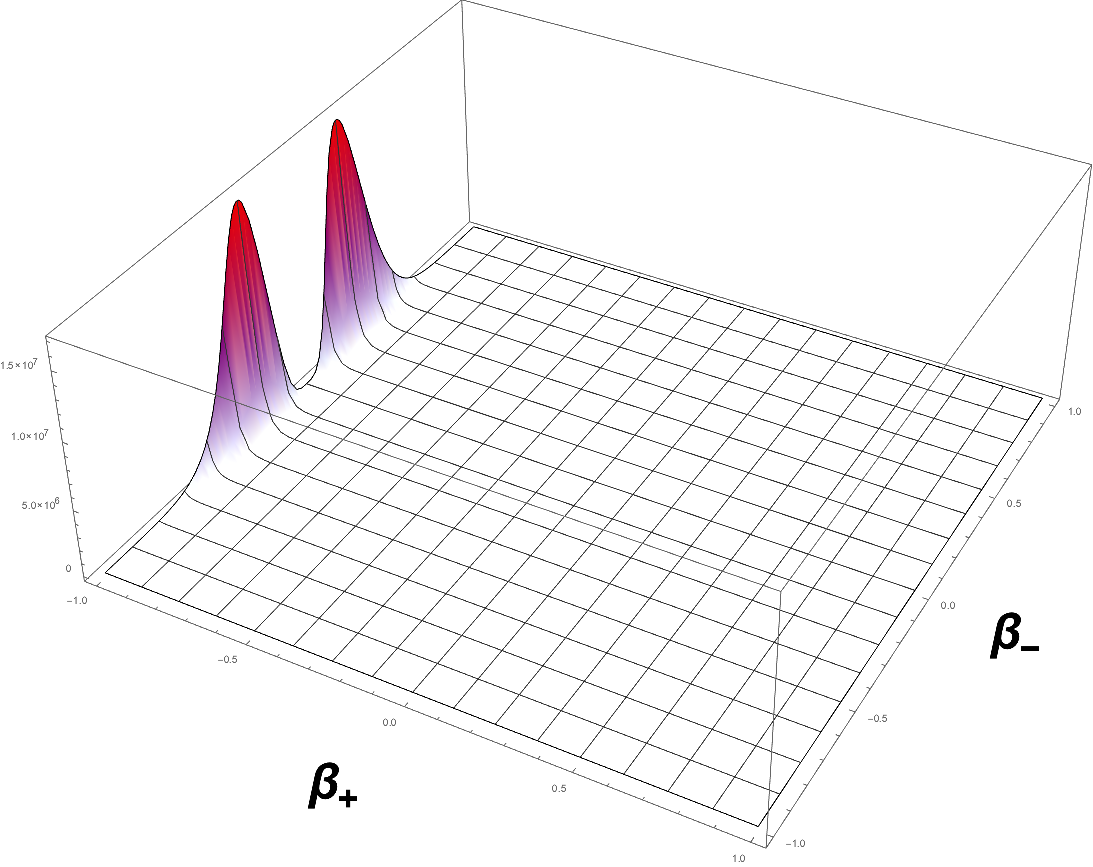}
\caption{$\alpha=0$}
\end{minipage}%
\end{figure}

For $\alpha=-3$ we see two localized maximums while for $\alpha=0$, we see no localized extremums. Rather the wave function is increasing without bound as $\beta_+$ approaches negative infinity. Because the goal of this paper is to solely report solutions, whether exact or approximate to the Bianchi VIII Wheeler DeWitt equation, we will leave the interpretation of these exotic looking wave functions for a future work.

\section{\label{sec:level1}Leading Order Solutions With Matter sources ($\Lambda \ne 0$ And Primordial Magnetic Field) And Without Matter}
The author has found six solutions to the following Bianchi VIII Euclidean-signature Hamilton Jacobi equation 

\begin{equation}
\begin{aligned} 
&{\left(\frac{\partial \mathcal{S}_{(0)}}{\partial \alpha}\right)^{2}-\left(\frac{\partial \mathcal{S}_{(0)}}{\partial \beta_{+}}\right)^{2}-\left(\frac{\partial \mathcal{S}_{(0)}}{\partial \beta_{-}}\right)^{2}}+U=0 \\ &
U=\left(f\right) 4 e^{6 \beta_+} \left(e^{6 \beta_+} \sinh ^2\left(2 \sqrt{3}
   \beta_-\right)+\cosh \left(2 \sqrt{3} \beta_-\right)\right) \\& +\frac{2 e^{6 a} \Lambda}{9 \pi }+2b^{2}e^{2\alpha-4\beta_+}+f\\& f=\frac{1}{3} e^{4 \alpha-8 \beta_+},
\end{aligned}
\end{equation}
for the case when only a cosmological constant is present $(b=0)$, where $b^{2}$ is a measure of how strong the primordial magnetic field is. These solutions can be used to construct leading order solutions to the Lorentzian signature Bianchi VIII Wheeler Dewitt equation when a cosmological constant is present without any other matter sources. The six solutions are the following
\begin{equation}
\begin{aligned}
\mathcal{S}^1_{(\pm 0)}:=&\pm \Biggl(\frac{1}{6} e^{2 \alpha-4 \beta_+} \left(2 e^{6 \beta_+} \cosh \left(2 \sqrt{3} 
\beta_-\right)-1\right)\\&+\frac{\Lambda e^{4 \alpha+4 \beta_+}}{36 \pi }\Biggr)
\end{aligned}
\end{equation}
\begin{equation}
\begin{aligned}
\mathcal{S}^2_{(\pm 0)}:=&\pm \Biggl(\frac{1}{6} e^{2 \alpha-4 \beta_+} \left(2 e^{6 \beta_+} \cosh \left(2 \sqrt{3} 
\beta_-\right)-1\right)\\&-\frac{\Lambda
 e^{4 \alpha+2 \sqrt{3} \beta_--2 \beta_+}}{36 \pi }\Biggr)
\end{aligned}
\end{equation}
\begin{equation}
\begin{aligned}
\mathcal{S}^3_{(\pm 0)}:=&\pm \Biggl(\frac{1}{6} e^{2 \alpha-4 \beta_+} \left(2 e^{6 \beta_+} \cosh \left(2 \sqrt{3} 
\beta_-\right)-1\right)\\&-\frac{\Lambda
 e^{4 \alpha-2 \sqrt{3} \beta_--2 \beta_+}}{36 \pi }\Biggr).
\end{aligned}
\end{equation}

The above leading order terms possess the strange property that the sign of $\Lambda$ associated with each one of them does not identically impart the same behavior to the Wheeler DeWitt wave function $\stackrel{(0)}{\Psi}_{\hbar}=e^{-\frac{1}{X} \mathcal{S}_{(0)}-\mathcal{S}_{(1)}-\frac{X}{2 !} \mathcal{S}_{(2)}-\cdots}$. This can be seen by comparing the sign of the $\Lambda$ term in (49) to the sign of the $\Lambda$ terms in (50) and (51). 

When both a cosmological constant and a primordial magnetic field are present the author has found the following two solutions

\begin{equation}
\begin{aligned}
\mathcal{S}^4_{(\pm 0)}:=&\pm \Biggl(\frac{1}{6} e^{2 \alpha-4 \beta_+} \left(2 e^{6 \beta_+} \cosh \left(2 \sqrt{3}
   \beta_-\right)-1\right)\\&  + \text{b}^2 (\alpha+\beta_+)+\frac{\Lambda e^{4 \alpha+4 \beta_+}}{36 \pi }\Biggr).
\end{aligned}
\end{equation}

The primordial nature of our magnetic field is captured by the fact that the term corresponding to it is only proportional to $\alpha$. As a result for $\alpha << 0$ the magnetic field dominates over the cosmological constant term and the anisotropic vacuum term, however for $\alpha >> 0$ the effects of the primordial magnetic field become negligible. It is always reassuring when certain classical properties of a theory manifest themselves in their quantum analogue.

Besides computing new leading order terms to the Wheeler DeWitt equation when a matter source is present, the author has found additional leading order terms when $\Lambda = 0$ by inserting this ansatz \begin{equation}
\begin{aligned}
S_{(0)}&=\frac{1}{6} \mathrm{e}^{2 \alpha+2 \beta_{+}}\left(-\mathrm{e}^{-6 \beta_{+}}+2 \cosh 2 \sqrt{3} \beta_{-}\right)\\&+\frac{1}{3} \mathrm{e}^{2 \alpha}\left[b \mathrm{e}^{-\beta_{+}+\sqrt{3} \beta_{-}}+d \mathrm{e}^{-\beta_{+}-\sqrt{3} \beta_{-}}+c \mathrm{e}^{2 \beta_{+}}\right]
\end{aligned}
\end{equation}
into the Bianchi VIII Euclidean-signature Hamilton Jacobi equation. The author obtained this ansatz by modifying the following ansatz which was originally\cite{barbero1996minisuperspace} developed to find solutions to the Bianchi IX Einstein-Hamilton-Jacobi equation
\begin{equation}
\begin{aligned}
S_{(0)}&=\frac{1}{6} \mathrm{e}^{2 \alpha+2 \beta_{+}}\left(\mathrm{e}^{-6 \beta_{+}}+2 \cosh 2 \sqrt{3} \beta_{-}\right)\\&+\frac{1}{3} \mathrm{e}^{2 \alpha}\left[b \mathrm{e}^{-\beta_{+}+\sqrt{3} \beta_{-}}+d \mathrm{e}^{-\beta_{+}-\sqrt{3} \beta_{-}}+c \mathrm{e}^{2 \beta_{+}}\right]
\end{aligned}
\end{equation}.
Inserting our ansatz (53) into the Bianchi VIII Euclidean-signature Wheeler DeWitt equation yields 
\begin{equation}
e^{3 \beta_+-\sqrt{3} \beta_-} (\text{b}+\text{c} \text{d})+e^{\sqrt{3} 
\beta_-+3 \beta_+} (\text{b} \text{c}+\text{d})+\text{b} \text{d}-\text{c} =0,
\end{equation}
if we solve for $\left(b,c,d\right)$ this results in the following 8 leading order terms
\begin{equation}
\begin{aligned}
S^{5}_{(0)}&=\pm \Biggl(\frac{1}{6} \mathrm{e}^{2 \alpha+2 \beta_{+}}\left(-\mathrm{e}^{-6 \beta_{+}}+2 \cosh 2 \sqrt{3} \beta_{-}\right)\\&+\frac{1}{3} \mathrm{e}^{2 \alpha}\left[-i \mathrm{e}^{-\beta_{+}+\sqrt{3} \beta_{-}}-i \mathrm{e}^{-\beta_{+}-\sqrt{3} \beta_{-}}- \mathrm{e}^{2 \beta_{+}}\right]\Biggr)
\end{aligned}
\end{equation}
\begin{equation}
\begin{aligned}
S^{6}_{(0)}&=\pm \Biggl(\frac{1}{6} \mathrm{e}^{2 \alpha+2 \beta_{+}}\left(-\mathrm{e}^{-6 \beta_{+}}+2 \cosh 2 \sqrt{3} \beta_{-}\right)\\&+\frac{1}{3} \mathrm{e}^{2 \alpha}\left[-i \mathrm{e}^{-\beta_{+}+\sqrt{3} \beta_{-}}+i \mathrm{e}^{-\beta_{+}-\sqrt{3} \beta_{-}}+ \mathrm{e}^{2 \beta_{+}}\right]\Biggr)
\end{aligned}
\end{equation}
\begin{equation}
\begin{aligned}
S^{7}_{(0)}&=\pm \Biggl(\frac{1}{6} \mathrm{e}^{2 \alpha+2 \beta_{+}}\left(-\mathrm{e}^{-6 \beta_{+}}+2 \cosh 2 \sqrt{3} \beta_{-}\right)\\&+\frac{1}{3} \mathrm{e}^{2 \alpha}\left[i \mathrm{e}^{-\beta_{+}+\sqrt{3} \beta_{-}}+i \mathrm{e}^{-\beta_{+}-\sqrt{3} \beta_{-}}- \mathrm{e}^{2 \beta_{+}}\right]\Biggr)
\end{aligned}
\end{equation}
\begin{equation}
\begin{aligned}
S^{8}_{(0)}&=\pm \Biggl(\frac{1}{6} \mathrm{e}^{2 \alpha+2 \beta_{+}}\left(-\mathrm{e}^{-6 \beta_{+}}+2 \cosh 2 \sqrt{3} \beta_{-}\right)\\&+\frac{1}{3} \mathrm{e}^{2 \alpha}\left[i \mathrm{e}^{-\beta_{+}+\sqrt{3} \beta_{-}}-i \mathrm{e}^{-\beta_{+}-\sqrt{3} \beta_{-}}+ \mathrm{e}^{2 \beta_{+}}\right]\Biggr) 
\end{aligned}
\end{equation}
Solving for the quantum corrections associated with the leading order terms we found in this section is a formidable task. As a result we will be content to study the simpler problem of finding quantum corrections, and conserved quantities which can be our $\phi_{0}$ for a select few of our leading orders terms when we restrict our wave functions to the $\beta_+$ axis. 

\section{\label{sec:level1}Bianchi VIII Wave Functions Restricted To The $\beta_+$ Axis}

We will now explore the quantum diagonalized Bianchi VIII models when our wave functions are restricted to the $\beta_+$ axis.  In other words we will explore the quantum analogue of classical states whose flow in minisuperspace initially begins on the $\beta_+$ axis and remains solely on that axis for all values of our evolution parameter $\tau$.

One can easily convince oneself that (52), (56), and (58) permit flows in minisuperpsace which originate on the $\beta_+$ axis, and remain on that axis for all values of our evolution parameter by following the straightforward analysis performed in section V of \cite{berkowitz2019new}. Assuming this as a given, it is important to keep in mind that despite studying wave functions whose amplitudes vanish on the $\beta_-$ axis, we are not studying the quantum LRS Bianchi VIII models. The author has compiled a large amount of preliminary results for that model and will be publishing them in due time. Even though we will be setting $\beta_-=0$ for some parts of our calculation, as can be seen by referring to \cite{berkowitz2019new} section V the sheer existence of a $\beta_-$ axis will have a noticeable effect on our solutions.

To begin our analysis we will first insert the entirety of (56) into equation (17), and take the derivatives with respect to $\alpha$, $\beta_+$, and $\beta_-$. Afterwards though we will set $\beta_-=0$ because even though we are only interested in wave functions which exist on the $\beta_+$ axis, the contribution from the derivatives of our leading order term with respect to $\beta_-$ does not vanish when $\beta_-=0$. Furthermore this means we cannot compute a $\mathcal{S}_{(2)}$ term using this method, because in order to do so we would need the full solution of the $\mathcal{S}_{(1)}$ equation for (56) so we can compute the value of its derivatives with respect to $\beta_-$ and set $\beta_-=0$.

The resulting equation we obtain from inserting (56) into (17) and setting $\beta_-=0$ is  

\begin{equation}
\begin{aligned}
&4 i e^{3 \beta_+} (2 \frac{\partial \mathcal{S}_{(1)}}{\partial \alpha}+\frac{\partial \mathcal{S}_{(1)}}{\partial \beta_+}+\text{B})-2 \frac{\partial \mathcal{S}_{(1)}}{\partial \alpha}-4 \frac{\partial \mathcal{S}_{(1)}}{\partial \beta_+} \\&+12 e^{6 \beta_+}-\text{B}-6=0.
\end{aligned}
\end{equation}

Because of its independence from $\alpha$ we can choose our solution to have the form 

\begin{equation}
\begin{aligned}
\mathcal{S}^{5}_{(1)}=x1\alpha+f(B)
\end{aligned}
\end{equation}

where x1 can be any real or complex number. This results in a differential equation that can be easily solved in Mathematica and gives the following as our $\mathcal{S}_{(1)}$

\begin{equation}
\begin{aligned}
\mathcal{S}^{5}_{(1)}=& \alpha \text{x1}+\frac{1}{8} \bigl(-2 \beta_+ (\text{B}+2 \text{x1}+6) \\&-\log \left(e^{6 \beta_+}+1\right)
   (\text{B}+2 \text{x1}-6) \\&+2 i \tan ^{-1}\left(e^{3 \beta_+}\right) (\text{B}+2
   \text{x1}-6)+8 i e^{3 \beta_+}\bigr).
\end{aligned}
\end{equation}

Inserting the entirety of (56) into (23), and following the same steps results in 

\begin{equation}
\begin{aligned}
\phi^{5}_{(0 )}=\left(\frac{e^{4 \alpha-2 \beta_+}}{\left(e^{3 \beta_+}+i\right)^2}\right)^{\text{c1}}
\end{aligned}
\end{equation}
where c1 can be any real or complex number. However in the interest of interpreting c1 as a graviton excitation number\cite{bae2014quantizing} we will only consider real values of c1. To avoid obtaining any pathological solutions whose interpretation are difficult to ascertain from simply looking at their wave functions we will use the $\boldsymbol{minus}$ signed form of (56) and have $c1 \geq 0 $. Our $\phi_{(0)}$ (63) does not vanish for any real values of $\alpha$ or $\beta_+$. Thus 'excited' states constructed from it are scattering states. Because the Wheeler DeWitt equation is linear, and we have  scattering states we can take the following superposition $\int_{0}^{\infty} dc1 e^{-c1^{2}} \phi^{5}_{(0)}e^{-\mathcal{S}^{5}_{(0 \hspace{1mm} \beta_-=0) }-\mathcal{S}^{5}_{(1)}}$. Two plots for our quantum Bianchi VIII scattering 'excited' state can be found in figures 3 and 4. 

\begin{figure}[!ht]
\begin{minipage}[c]{0.4\linewidth}
\includegraphics[scale=.12]{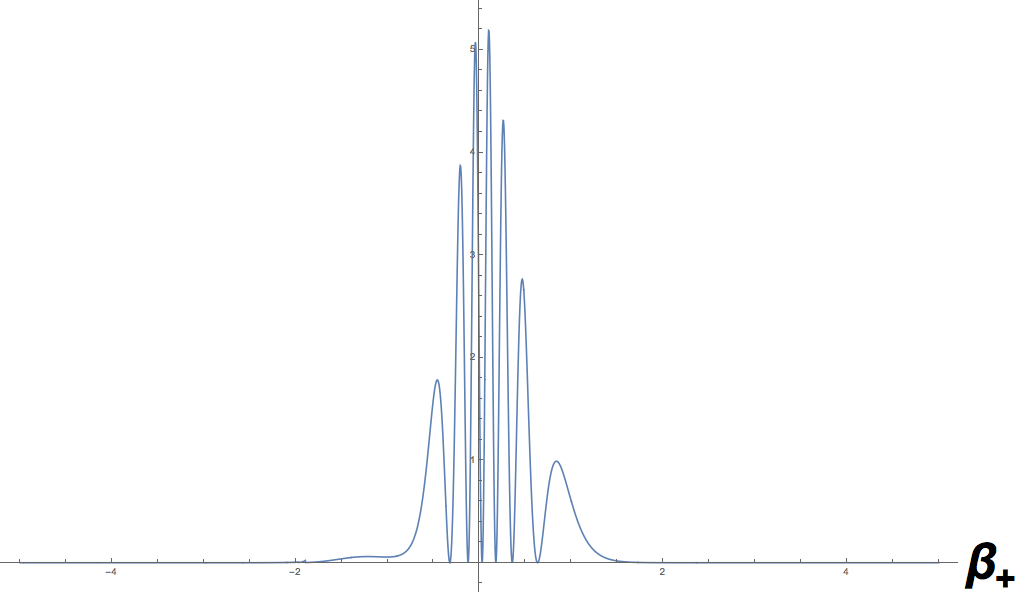}
\caption{$\alpha=-4$}
\end{minipage}

\begin{minipage}[c]{0.4\linewidth}
\includegraphics[scale=.12]{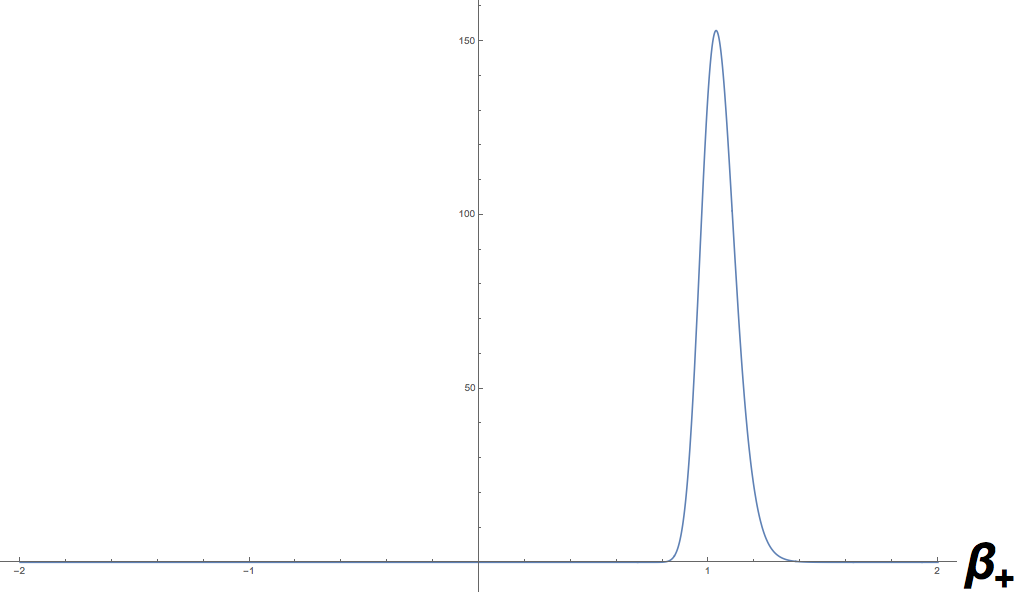}
\caption{$\alpha=4$}
\end{minipage}%
\end{figure}

Using $\alpha$ as our clock, this particular superposition of  scattering states forms many peaks around values of $\beta_+$ for $\alpha << 0$. Physically this represents a Bianchi VIII universe which can tunnel between different sharply defined states of anisotropy determined by $\beta_+$. However when it grows larger such as when $\alpha >>0$ the wave function of the Bianchi VIII universe becomes sharply peaked at a particular value of $\beta_+$. Physically this picture makes sense, tunneling is mainly a quantum phenomenon, thus it should strongly be associated with a universe which is spatially very small or hot and dense. The more negative $\alpha$ is the smaller the scale factor of our Bianchi VIII universe is. Universes which have a large scale factor or $\alpha >>0$ should be classical in nature, and thus have wave functions which are sharply peaked. We should stress that we are only dealing with approximate solutions with only a single quantum correction, but nonetheless these results are somewhat expected and quite interesting at the same time. 

A subtle feature of our superposition is that for small $\alpha << 0$ our wave function as seen in figure 3 aesthetically has the characteristics one would expect of an 'excited' state; while the wave function for $\alpha >> 0$ has the characteristics of a 'ground' state. As we mentioned when $\alpha << 0$ the universe described by the wave function in figure 3 is quantum in nature because it has a high likelihood of tunneling into another geometric configuration, while the wave function in figure 4 describes a universe classical in nature where quantum phenomena such as tunneling is exceedingly unlikely. Thus it is possible that the excited/ground state dichotomy that is well defined in ordinary quantum mechanics is not the best approach to delineate states in Wheeler DeWitt quantum cosmology because states that behave as 'excited' states for a certain range of $\alpha$ can also behave as 'ground' states in another range of $\alpha$. Perhaps a more useful way to distinguish states is by whether or not they permit a quantum mechanical universe where tunneling is a common phenomenon for some range of the parameters that the state depends on. The question of how we distinguish the eigenstates of an operator with vanishing eigenvalues $\hat{\mathcal{H}}_{\perp} \Psi =0$ is one that deserves to be investigated further. 

We will skip applying the same analysis to (58) because the wave function we would obtain behaves very similarly to the one we just studied. For (52) we will only study the semi-classical limit. If we insert (52) into equation (23) and solve the PDE which results when $\beta_-$ is set to zero one obtains the following $\phi_{0}$

\begin{equation}
\begin{aligned}
\phi^{4}_{(0)}=& \bigl(e^{2 \alpha-2 \beta_+} \bigl(\Lambda e^{4 \alpha+8 \beta_+}  +9 \pi  e^{2 \alpha+6 \beta_+}\\&+9 \pi  e^{2 \alpha}+27 \text{b}^2 \pi  e^{4 
\beta_+}\bigr)\bigr)^{\text{c1}}.
\end{aligned}
\end{equation}

Because $b^{2}$ is always positive our 'excited' states obtained from this family of conserved quantities are bound states when  $\Lambda < 0$ and scattering states when $\Lambda > 0$. If we plot $\phi^{4}_{(0)}e^{-\mathcal{S}_{(0 \hspace{1mm} 52 \hspace{1mm} \beta_-=0)}}$ where our semi-classical term is the minus sign variation of (52) we obtain the following two interesting pictures. 

\begin{figure}[!ht]
\begin{minipage}[c]{0.4\linewidth}
\includegraphics[scale=.09]{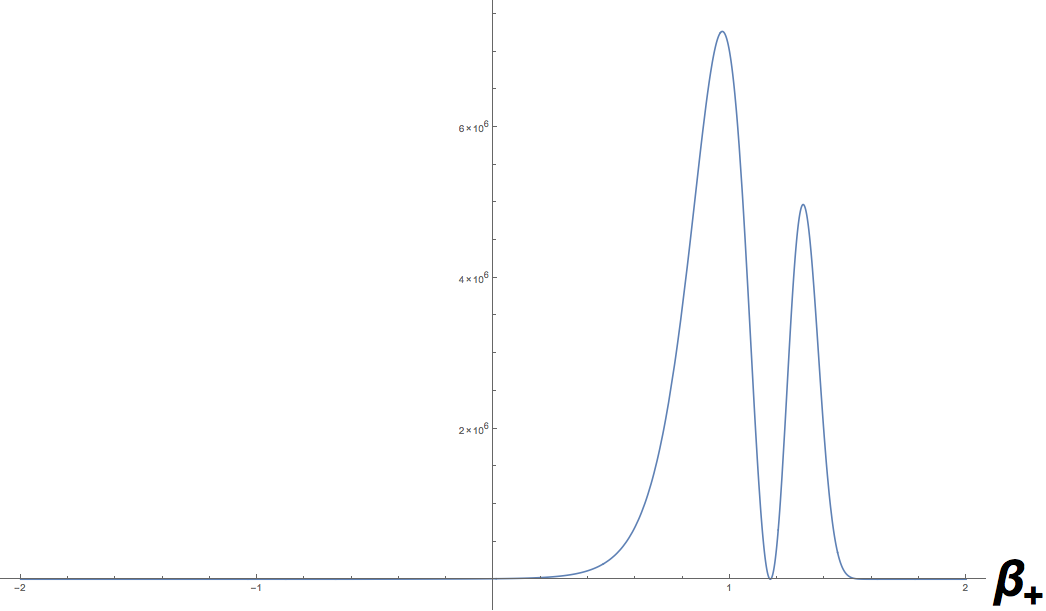}
\caption{$\alpha=\frac{1}{2}$,  \\ $\Lambda=-1$,\hspace{1mm} b=0,\hspace{1mm} c1=2}
\end{minipage}
\hfill
\begin{minipage}[c]{0.4\linewidth}
\includegraphics[scale=.09]{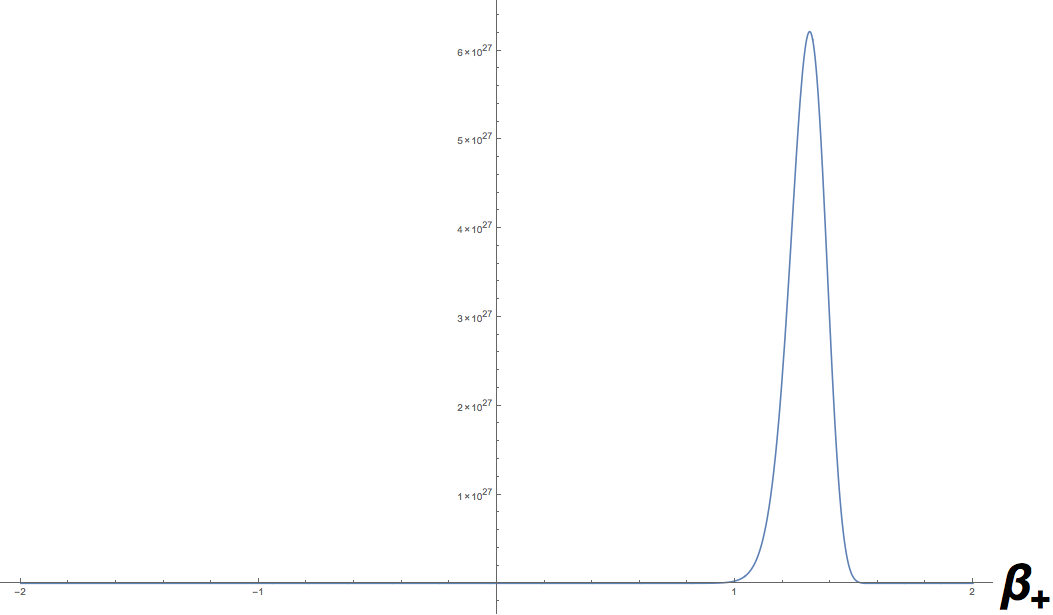}
\caption{$\alpha=\frac{1}{2}$,\\ $\Lambda=-1$,\hspace{1mm} b=5,\hspace{1mm} c1=2}
\end{minipage}%
\end{figure}

When the strength of our magnetic field is zero the Bianchi VIII universe described by this wave function for $\alpha=\frac{1}{2}$ can tunnel between two sharply peaked geometric configurations. However, when the primordial magnetic field is turned on it causes this particular Bianchi VIII universe to peak sharply at a particular value of $\beta_+$. A stronger primordial magnetic field results in an universe with a less "fuzzy", or in other words a more definitive geometric configuration. These results for a Bianchi VIII universe with a cosmological constant and a primordial magnetic field restricted to the $\beta_+$ axis are similar to what the author obtained when he studied the diagonalized quantum Bianchi IX models under the same circumstances.

\section{\label{sec:level1}Final Remarks}
The quantum cosmology of the Bianchi VIII models have been seldom studied compared to the quantum cosmology of the Bianchi IX models\cite{misner1969quantum,misner1969mixmaster,klauder1972magic,graham1993supersymmetric,bae2015mixmaster,graham1991supersymmetric,berkowitz2019new}; however recent work has been undertaken in studying the quantum cosmology of the Bianchi VIII LRS model \cite{karagiorgos2019quantum}. As previously mentioned the author himself has amassed preliminary results for the quantum Bianchi VIII LRS model which he will publish in due time, including solutions for arbitrary Hartle Hawking ordering parameter when stiff matter is present. Furthermore, using the Euclidean-signature semi classical he has amassed results for Bianchi VII, II, and I models with a cosmological constant which will also be published in due time.

In this paper, we expanded the literature on the quantum cosmology of the diagonalized Bianchi VIII models by finding new closed form solutions to its corresponding Wheeler DeWitt equation for Hartle Hawking ordering parameters $B=\pm 2\sqrt{33}$ using the Euclidean-signature semi classical method. In addition, we were able to define 'ground' states and 'excited' states for the quantum Bianchi VIII models and show that these 'excited' states are a hybrid of scattering states and bound states. We also found a plethora of solutions for the Bianchi VIII Euclidean-signature Hamilton Jacobi equation for the case when $\Lambda \ne 0$ and an aligned primordial magnetic field is present, and when $\Lambda = 0$ which on their own can be used to construct leading order/semi-classical solutions to the Lorentzian-signature Bianchi VIII symmetry reduced Wheeler Dewitt equation. We then obtained some interesting plots of scattering states when we restricted the study of our wave function to the $\beta_+$ axis. The results in this paper greatly expand our ability to analyze the quantum cosmology of the diagonalized Bianchi VIII model and the author greatly looks forward to see what physics can be extrapolated from the mathematical results in this paper. Furthermore the potential applications of the Euclidean-signature semi classical method to problems in quantum cosmology and beyond are quite vast. As a result the author very much looks forward to seeing what future applications of this method will produce.

\section{\label{sec:level1}ACKNOWLEDGMENTS}
 
I am grateful to Professor Vincent Moncrief for valuable discussions at every stage of this work. I would also like to thank George Fleming for facilitating my ongoing research in quantum cosmology. Daniel Berkowitz acknowledges support from the United States Department of Energy through grant number DE-SC0019061. I also must thank my aforementioned parents.

\bibliography{BianchiV}

\end{document}